\documentclass[11pt]{article}
\usepackage{graphicx}
\def \d {{\rm d}}

\def \U {{\cal U}}
\def \V {{\cal V}}

\begin{document}

\title{A snapping cosmic string in a de~Sitter\\ or anti-de~Sitter
universe}

\author{J. Podolsk\'y\thanks{E--mail: {\tt Podolsky@mbox.troja.mff.cuni.cz}}
\\
  Institute of Theoretical Physics, Charles University in Prague,\\
V Hole\v{s}ovi\v{c}k\'ach 2, 18000 Prague 8, Czech Republic.\\
\\
and J. B. Griffiths\thanks{E--mail: {\tt J.B.Griffiths@Lboro.ac.uk}} \\
Department of Mathematical Sciences, Loughborough University \\
Loughborough, Leics. LE11 3TU, U.K. \\ }

\maketitle

\begin{abstract}
\noindent
We present and describe an exact solution of Einstein's equations which
represents a snapping cosmic string in a vacuum background with a
cosmological constant~$\Lambda$. The snapping of the string generates an
impulsive spherical gravitational wave which is a particular member of a
known family of such waves. The global solution for all values of
$\Lambda$ is presented in various metric forms and interpreted
geometrically. It is shown to represent the limit of a family of sandwich
type~N Robinson--Trautman waves. It is also derived as a limit of the
C-metric with $\Lambda$, in which the acceleration of the pair of black
holes becomes unbounded while their masses are scaled to zero.
\end{abstract}

\section{Introduction}

An exact solution of Einstein's equations is known \cite{GlePul89},
\cite{Bicak90} which describes an impulsive spherical gravitational wave
in a Minkowski background that is generated by a snapping infinite cosmic
string (or by an expanding string inside the sphere \cite{BicSch89}). This
solution can be obtained either \cite{GlePul89} by pasting two appropriate
forms of Minkowski space either side of the spherical wavefront, or
\cite{Bicak90} as a limiting case of particular solutions with
boost-rotation symmetry (see also \cite{BicSch89}--\cite{PodGri01b}). In
both constructions, the two ends of the semi-infinite strings recede from a
common point generating an impulsive spherical wave. (Alternatively,
there is an expanding cosmic string along the axis of symmetry separating
the two particles.) However, as already pointed out \cite{Bicak90}, this
solution does not strictly describe a snapping cosmic string, but rather
two semi-infinite cosmic strings which initially approach at the speed of
light and then separate again at the instant at which they collide.
The purpose of the present paper is to present and investigate an analogous
solution which describes a snapping cosmic string in a de~Sitter or
anti-de~Sitter universe, thus generating an impulsive spherical wave in
either of these backgrounds.

In fact, a general method for constructing expanding impulsive spherical
gravitational waves in a Minkowski background had previously been suggested
by Penrose \cite{Pen72}. This involves cutting Minkowski space-time along
a null cone and then re-attaching the two pieces with a suitable ``warp''.
The explicit general solution written in a continuous coordinate system is
given in \cite{NutPen92}--\cite{AliNut01}. This, of course, includes the
solution which describes a snapping cosmic string as a particular case.

This general family of solutions has also been extended to include a
non-zero cosmological constant $\Lambda$ \cite{Hogan92}, \cite{PodGri00}.
These solutions describe a family of expanding spherical gravitational
waves in a de~Sitter or anti-de~Sitter universe. In this paper, we identify
the specific solution which represents a snapping cosmic string in these
backgrounds. Before presenting this, however, we first describe an
instructive geometrical representation of a cosmic string in the 
(anti-)de~Sitter background.

Further, the general class of solutions of this type has been shown
\cite{PodGri99} to be equivalent to impulsive limits of the class of
Robinson--Trautman type~N solutions. In section~4, we will present the
exact solution for a snapping string in this context. In section~5, we
will demonstrate how this solution can also be obtained as a limit of the
C-metric which can be interpreted as the field of two black holes
accelerating away from each other under the action of strings which pull
in opposite directions. In the limit as the acceleration $A$ becomes
unbounded while the mass $m$ of the black holes is reduced to zero keeping
the quantity $mA$ constant, this reduces exactly to the solution for a
snapping cosmic string. For a Minkowski background, this procedure is
known~\cite{PodGri01b}. However, the method applied there cannot be
extended to the case when the cosmological constant is non-zero as no
global coordinate system (analogous to that for the Weyl coordinates) is
known for this case. A different method is therefore developed in
section~5 in which the solution for a snapping cosmic string in a
de~Sitter or anti-de~Sitter background can be derived as the limit of the
generalization of the C-metric which contains a cosmological constant.

\section{A 5-dimensional representation}

As is well known, the de~Sitter and anti-de~Sitter space-times can be
represented as the 4-dimensional hyperboloid
  \begin{equation}
   {Z_0}^2-{Z_1}^2-{Z_{23}}^2-\varepsilon{Z_4}^2= -\varepsilon a^2,
  \label{hyp}
  \end{equation}
  in the flat 5-dimensional space
  \begin{equation}
  \d s^2=\d{Z_0}^2-\d{Z_1}^2-\d{Z_{23}}^2 -{Z_{23}}^2\,e^{-2c}\,\d\Phi^2
  -\varepsilon\d{Z_4}^2,
  \label{dSin5D}
  \end{equation}
  where $a^2={3/|\Lambda|}$, $\varepsilon$ is the sign of $\Lambda$, $c$ is
a constant, and $Z_{23}\in[0,\infty)$ and $\Phi\in[0,2\pi)$ are related to
standard cartesian coordinates by $Z_2=Z_{23}\cos\Phi$ and
$Z_3=Z_{23}\sin\Phi$. When $c=0$ and $\Lambda>0$, this is a complete
de~Sitter space in which, at any time $Z_0=$ const., the universe is
closed and has the geometry of a 3-sphere. When $c>0$, a wedge in the
space-time is missing, and the de~Sitter space may be considered to
include a string with a deficit angle $2\pi(1-e^{-c})$ located along the
line $Z_{23}=0$. It may immediately be observed from (\ref{hyp}) that the
string lies on a closed meridian around the universe. As the universe
contracts (for $Z_0<0$) and then expands (for $Z_0>0$), the length of the
string also correspondingly contracts and expands.

Analogously, when $c=0$ and $\Lambda<0$, this is the complete open
anti-de~Sitter space. When $c>0$, a wedge in the space-time is
missing, and the space may be considered to include an infinite
string with a deficit angle $2\pi(1-e^{-c})$ located along the line
$Z_{23}=0$.

\begin{figure}[hpt]
\begin{center} \includegraphics[scale=0.6, trim=5 5 5 -5]{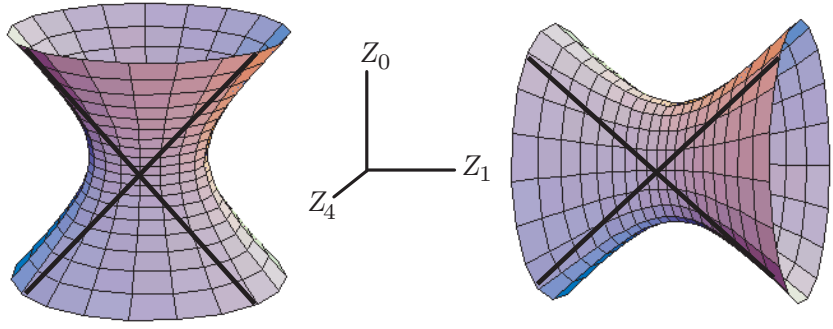}
\caption{\small The hyperboloids represent the de~Sitter and anti-de~Sitter
backgrounds with $Z_{23}=0$. The expanding spherical impulsive waves are
located on the sections with $Z_4=a$. In these sections, strings are
located in the region with $Z_4<a$ in the de~Sitter background and $Z_4>a$
in the anti-de~Sitter background. }
\end{center}
\end{figure}

It will be shown in the following sections that the ends of a snapped
cosmic string in both the de~Sitter and the anti-de~Sitter universe
propagate along the null generators of the hyperboloid which are located
on the section $Z_4=a$ which corresponds to the poles of an expanding
sphere on which ${Z_1}^2+{Z_{23}}^2={Z_0}^2$. This is illustrated in
figure~1. In the de~Sitter case, the string is contained in the region for
which $Z_4<a$, but is absent for $Z_4>a$. However, for $\Lambda<0$, the
string is contained in the region for which $Z_4>a$, and is absent for
$Z_4<a$. For $Z_4\ne a$, such a situation can be formally represented by
the line element
  \begin{equation}
  {\d s_0}^2=\d{Z_0}^2-\d{Z_1}^2-\d{Z_{23}}^2
 -{Z_{23}}^2\,e^{-2c\Theta(\varepsilon(a-Z_4))}\,\d\Phi^2
 -\varepsilon\d{Z_4}^2,
  \label{discont}
  \end{equation}
  where $\Theta$ is the Heaviside step function. However, the metric
(\ref{discont}) does not correspond to a global solution of Einstein's
equations. The discontinuity across \ $Z_4=a$ \ must be accompanied by an
impulsive component which can be interpreted as an expanding spherical
gravitational wave. The complete exact solution will be presented below.

\section{An explicit solution in a continuous metric form}

We will now use the Penrose ``cut and paste'' method \cite{Pen72} (see also
\cite{NutPen92}, \cite{PodGri00} and \cite{PodGri99}) to construct an
explicit solution representing a snapping cosmic string in a background
with a non-zero cosmological constant. It is convenient to represent the
(anti-)de~Sitter background by a line element in the form
  \begin{equation}
  \d s^2 = {2\,\d\U\,\d\V -2\,\d\eta\,\d\bar\eta \over
\big[1-{1\over6}\Lambda(\U\V-\eta\bar\eta)\big]^2} ,
  \label{metric1}
  \end{equation}
  which is related to the parameterization of the hyperboloid
(\ref{hyp}) by the substitution
  $$ \U=\sqrt2\,a\,{Z_0+Z_1\over Z_4+a}, \qquad
\V=\sqrt2\,a\,{Z_0-Z_1\over Z_4+a}, \qquad
\eta=\sqrt2\,a\,{Z_{23}\exp(i\Phi)\over Z_4+a}. $$
  Starting with the line element (\ref{metric1}), we first perform the
transformation
  \begin{equation}
  \U= U+V\,Z\bar Z/p, \qquad
  \V= V/p+\epsilon U, \qquad
  \eta= V\,Z/p,
  \label{inv}
  \end{equation}
  where
  $$ p=1+\epsilon Z\bar Z, \qquad \epsilon=-1,0,1. $$
  (It can be seen from \cite{GrPoDo02} that the most natural choice of
parameter for these solutions is $\epsilon=1$. However, in the impulsive
limit, all choices are permitted and the parameter $\epsilon$ is retained
below.)  With the transformation (\ref{inv}), (\ref{metric1}) becomes
  \begin{equation}
\d s^2 = {{ \displaystyle 2\d U\,\d V  +2\epsilon\,\d U^2
  -2 (V/p)^2\,\d Z\,\d\bar Z }\over
\big[1-{1\over6}\Lambda U(V+\epsilon U)\big]^2}.
  \label{U<0}
  \end{equation}
  The alternative transformation
  \begin{eqnarray}
  \V&=& {|Z|^{\delta-2}\over1-\delta} \left({|Z|^2\over p}\>V
+\left[\left({\delta\over2}\right)^2
+\left(1-{\delta\over2}\right)^2\epsilon|Z|^2\right]\>U\right),
\nonumber\\
  \U&=& {|Z|^{-\delta}\over1-\delta} \left({|Z|^2\over p}\>V
+\left[\left(1-{\delta\over2}\right)^2
+\left({\delta\over2}\right)^2\epsilon|Z|^2\right]\>U\right),
  \label{transe3} \\
  \eta&=& {Z^{1-\delta}|Z|^{\delta-2}\over1-\delta} \left({|Z|^2\over p}\>V
+\left({\delta\over2}\right)
\left(1-{\delta\over2}\right)p\>U\right),  \nonumber
  \end{eqnarray}
  where $\delta\in[0,1)$ is a constant parameter, puts the metric
(\ref{metric1}) into the form
  \begin{equation}
\d s^2 = {{ \displaystyle 2\,\d U\,\d V +2\epsilon\,\d U^2
-2\left| (V/p)\,\d Z-U\,p\,\bar H\,\d\bar Z \right|^2} \over
\big[1-{1\over6}\Lambda U(V+\epsilon U)\big]^2} ,
  \label{U>0}
  \end{equation}
  where
  \begin{equation}
  H={{\delta\over2}\big(1-{\delta\over2}\big)\over Z^2}.
  \label{H1}
  \end{equation}
 Apart from some trivial changes of notation, this corresponds to the
general transformation given in \cite{PodGri00} with the specific choice
$h(Z)=Z^{1-\delta}$ which, when $\delta\ne0$, introduces a cosmic string
with deficit angle $2\pi\delta$.

We now combine the above line elements by attaching (\ref{U<0}) for $U>0$
with (\ref{U>0}) for $U<0$. The resulting metric is expressed explicitly as
  \begin{equation}
\d s^2 = {{ \displaystyle 2\d U\,\d V +2\epsilon\,\d U^2
- 2\left| (V/p)\,\d Z-U\Theta(-U)\,p\,\bar H\,\d\bar Z \right|^2 }
\over \big[1-{1\over6}\Lambda U(V+\epsilon U)\big]^2}.
  \label{contmet}
  \end{equation} 
The line element (\ref{contmet}) is  continuous in $U$. However, there
is a discontinuity in the derivatives of the metric on the null
hypersurface $U=0$. This yields an impulsive component in the curvature
tensor component \ $\Psi_4=(p^2H/V)\,\delta(U)$ \ using a natural tetrad,
indicating the presence of an impulsive gravitational wave on this
hypersurface. In the above metrics, this hypersurface represents an
expanding sphere given by \ \hbox{$\U\V-\eta\bar\eta=0$}. \ In the
de~Sitter or anti-de~Sitter background, this corresponds to the section \
$Z_4=a$ \ in accordance with the geometrical description given in
section~2 (see figure~1). Moreover, since
  $$ U(V+\epsilon U) ={\cal U}{\cal V}-\eta\bar\eta
={6\over\Lambda}\,{(Z_4-a)\over(Z_4+a)}, $$
  for \ $V+\epsilon U>0$, \ it can be seen that the region inside the
spherical wavefront, for which $U>0$, corresponds to $Z_4>a$ in a
de~Sitter background, but to $Z_4<a$ in an anti-de~Sitter background.
Similarly, the outside region $U<0$ which contains the strings corresponds
to $Z_4<a$ when $\Lambda>0$ and to $Z_4>a$ when $\Lambda<0$.

In the combined metric (\ref{contmet}), the argument of the complex
coordinate $Z$ has the full range $\arg Z\in[0,2\pi)$ everywhere. In the
region inside the spherical wave in which $U>0$, it can be seen from
(\ref{inv}) that \ $\Phi=\arg\eta=\arg Z$ \ so that, for this region,
$\Phi$ has the same full range $[0,2\pi)$. Thus, inside the spherical wave,
the space-time is locally complete, corresponding to $c=0$ in
(\ref{dSin5D}). However, in the outer region where $U<0$, the expression
for $\eta$ in (\ref{transe3}) implies that \ $\Phi=\arg\eta=(1-\delta)\arg
Z$ \ so that here \ $\Phi\in[0,(1-\delta)2\pi)$. \ Thus, for
$\delta\in(0,1)$, a wedge of the space-time is missing. In order to
express this in the form (\ref{dSin5D}), it is necessary to rescale the
coordinate $\Phi$. This is equivalent to setting a non-zero value for $c$
such that \ $e^{-c}=1-\delta$, \ corresponding to the presence of a string
with deficit angle $2\pi\delta$ in the region $U<0$ outside the impulsive
wave.

\section{Limit of Robinson--Trautman sandwich waves}

It was argued in \cite{PodGri99} that the above class of solutions
for expanding impulsive spherical gravitational waves can be considered to
be impulsive limits of the family of vacuum Robinson--Trautman type~N
solutions with a cosmological constant. This has been further clarified in
\cite{GriDoc02} and \cite{GrPoDo02} using the standard form of the line
element
  \begin{equation}
  \d s^2 =2\,\d u\,\d r
+\Big[2\epsilon-2r(\log P)_u -{\Lambda\over3}r^2\Big]\d u^2
-2{r^2\over P^2}\,\d\zeta\,\d\bar\zeta,
  \label{RTmetric}
  \end{equation}
  where the function $P(\zeta,\bar\zeta,u)$ is given by
  $$ P=\big(1+\epsilon F\bar F\big)
\big(F_\zeta\bar F_{\bar\zeta}\big)^{-1/2}, $$
  in which $F=F(\zeta,u)$ is an arbitrary complex function of $u$ and
$\zeta$, holomorphic in $\zeta$. In \cite{GrPoDo02}, we considered the
particularly simple case in which
  \begin{equation}
  F(\zeta,u)=\zeta^{g(u)},
  \label{F}
  \end{equation}
  where $g(u)$ is an arbitrary positive function of retarded time. For this
choice, we obtain
   \begin{equation}
  P^2={\left[1+\epsilon(\zeta\bar\zeta)^g\right]^2
\over g^2(\zeta\bar\zeta)^{g-1}}\ , \qquad
  (\log P)_u =-{\>g'\over g} \left( 1 +{\textstyle{1\over2}}
\log(\zeta\bar\zeta)^g
\left[ {1-\epsilon(\zeta\bar\zeta)^g\over1+\epsilon(\zeta\bar\zeta)^g} \right]
   \right),
   \label{functions}
   \end{equation}
  so that the metric is obviously axially symmetric, and the only non-zero
component of the Weyl tensor using a natural tetrad is given by
   \begin{equation}
  \bar\Psi_4=-{1\over 2r}{\bar\zeta\over\zeta}
\left({1\over|\zeta|^{g}}+\epsilon|\zeta|^{g}\right)^2\ {g'\over g}.
   \label{Psi4case}
   \end{equation}
  For regions in which $g$ is a constant, this solution is conformally flat
and describes part of a Minkowski, de Sitter or anti-de Sitter background.
For a more general $g(u)$ it represents an exact gravitational wave with
amplitude given by (\ref{Psi4case}). On any wave surface $u=$~const.,
the complex number $F=\zeta^g$ represents a stereographic-type coordinate.
If we assume that the argument of $\zeta$ covers the full range
$[0,2\pi)$, then $\arg F\in[0,2\pi g)$ and these surfaces include a
deficit angle $2\pi(1-g)$ about the symmetry axis $F=0$.

It is therefore a simple matter to construct the above model for a snapping
cosmic string by setting $g=1$ for $u>0$, and $g=e^{-c}$ for $u<0$. This
can be expressed as
   \begin{equation}
  g(u)=e^{-c\,\Theta(-u)},
   \label{g}
   \end{equation}
  so that \ $g'/g=c\,\delta(u)$ \ and the Weyl tensor (\ref{Psi4case}) only
has an impulsive component on $u=0$. Although this produces a
discontinuity in the metric function $P$ and an impulsive component in
$(\log P)_u$, this construction may clearly be interpreted as the limit of
a ``well behaved'' sandwich wave of the Robinson--Trautman family in which
the metric is continuous.

The geometrical structure of these solutions and the character of the
singularities for different values of $\epsilon$ and $\Lambda$ have been
described in \cite{GrPoDo02} for any $g(u)$. Here, we apply this for an
impulsive spherical wave generated by a snapping string in a de~Sitter or
anti-de~Sitter background.

It is known that the metric (\ref{RTmetric}) can also be expressed in
terms of Garc\'{\i}a--Pleba\'nski coordinates \cite{GarPle81}, in the form
  \begin{eqnarray}
  \d s^2 &=& 2\,\d u\,\d r
+\Big(2\epsilon-{\Lambda\over3}r^2\Big)\d u^2 -2{r^2\over\psi^2}\,
\Big|\,\d\xi-f\,\d u\Big|^2
\nonumber\\
&&\  +r\,\Big[ (f_\xi+\bar f_{\bar\xi})
-{2\epsilon\over\psi}(\bar\xi f+\xi\bar f)
\Big] \d u^2, \nonumber
  \end{eqnarray}
  where $f(\xi,u)$ is an arbitrary holomorphic function and \
$\psi=1+\epsilon\xi\bar\xi$. \ This is achieved using the transformation \
$\xi=F(\zeta,u)$ \ such that \ $F_u=f(F(\zeta,u),u)$. \ With the above
specific choice (\ref{F}), this corresponds to
  \begin{equation}
  f(\xi,u)={g'\over g}\,\xi\log\xi.
  \label{f}
  \end{equation}
  (Notice that this form can be found in the list \cite{SaGaPl83} of
Robinson--Trautman solutions with symmetries corresponding to an axial
symmetry.) In particular, with (\ref{g}), which represents a snapping
cosmic string accompanied by an impulsive spherical gravitational wave,
the line element becomes
  \begin{eqnarray}
  \d s^2 &=& 2\,\d u\,\d r +\Big(2\epsilon-{\Lambda\over3}r^2\Big)\d u^2
-2{r^2\over(1+\epsilon\xi\bar\xi)^2}\,
\Big|\,\d\xi-c\,\xi\log\xi\,\delta(u)\,\d u\Big|^2
\nonumber\\
&&\qquad  +2c\,r\,\bigg[ 1+{1-\epsilon\xi\bar\xi\over1+\epsilon\xi\bar\xi}
\log|\xi| \bigg] \delta(u) \d u^2. \nonumber
  \end{eqnarray}
  Unfortunately, this form is mathematically problematic as it explicitly
contains a square of a delta function corresponding to the impulsive
gravitational wave localised on $u=0$.

\section{Null limit of the C-metric}

Let us now consider the well known C-metric which describes two black holes
which accelerate away from each other under the action of two semi-infinite
strings which pull them from infinity. It has been shown \cite{PodGri01b}
that the null limit of this solution (as the acceleration $A$ tends to
infinity while the mass $m$ tends to zero) is the solution for a
snapping cosmic string in a Minkowski background. The generalisation of
the C-metric to include a cosmological constant is well known. It is
therefore appropriate to investigate whether a solution for a snapping
cosmic string in a de~Sitter or anti-de~Sitter background can similarly be
constructed by taking the null limit of this solution. 
However, it is found that the methods used in a Minkowski background do
not extend to the case in which the cosmological constant is non-zero. For
this case, a more complicated procedure is required.

The C-metric with a cosmological constant can be expressed by the line
element
  \begin{equation}
  \d s^2 = {1\over A^2(x+y)^2} \left(F\>\d t^2 -{1\over F}\>\d y^2
  -G\>\d\phi^2 -{1\over G}\>\d x^2 \right),
  \label{Cmetric2}
  \end{equation}
  where
  \begin{equation}
  G(x) = 1 -x^2 - 2mA\,x^3, \qquad
  F(y) = -{\Lambda\over3A^2}-1 +y^2 - 2mA\,y^3.
  \label{FG}
  \end{equation}
  This is included in the general class of solutions given in
\cite{PleDem76}. A physical interpretation and analysis of these
space-times has been presented very recently in
\cite{PodGri01c}--\cite{PoOrKr03} where references to previous work can be
found.

In the above form of the metric, however, it is not possible to retain the
cosmological constant in the limit as $A\to\infty$ while $mA$ is kept
constant. Moreover, the limit in a Minkowski background was obtained
\cite{PodGri01b} using boost-rotation symmetric coordinates which are
related to the Weyl form of the metric and this does not exist when the
cosmological constant is non-zero. However, we can perform the
transformation
  $$ \left. \matrix{
{\displaystyle t+\int{\d y\over F}=Au} \cr
\noalign{\medskip}
{\displaystyle x+y={1\over Ar}} \cr
\noalign{\medskip}
{\displaystyle \phi = -i\,{\zeta-\bar\zeta\over\sqrt2}} \cr
\noalign{\medskip}
{\displaystyle \int{\d x\over G} =
  Au -{\zeta+\bar\zeta\over\sqrt2}} \cr
} \quad\right\}
  \qquad \Leftrightarrow \qquad
  \left\{ \quad \matrix{
{\displaystyle r={1\over A(x+y)}} \cr
\noalign{\medskip}
{\displaystyle u= {1\over A} \left( t+\int{\d y\over F} \right)} \cr
\noalign{\medskip}
\zeta= {\displaystyle {1\over\sqrt2} \left(  t+\int{\d y\over F}
-\int{\d x\over G}  +i\phi \right)} \cr
} \quad \right. $$
  which puts the metric (\ref{Cmetric2}) into the form of the
Robinson--Trautman solutions
  $$ \d s^2 =2\d u\d r +2H\d u^2 -2{r^2\over P^2}\d\zeta\d\bar\zeta, $$
  where
  $$ 2H(r,u,\zeta,\bar\zeta)= \Delta(\log P) -2r(\log P)_u
  -{2m\over r} -{\Lambda\over3}r^2, $$
  with \ $\Delta=2P^2\partial_\zeta\partial_{\bar\zeta}$, \ and with the
special expression for $P$ given by
  \begin{equation}
  P(\zeta,\bar\zeta,u)= G^{-1/2}
\big(x(\zeta,\bar\zeta,u)\big).
  \label{specialP}
  \end{equation}
  It can be shown, as required, that (\ref{specialP}) satisfies the
Robinson--Trautman (vacuum) equation.

This form of the C-metric has the immediate advantage that the dependence
on $A$ and $\Lambda$ is completely separated. The cosmological constant
$\Lambda$ only occurs linearly in the expression for~$H$, whereas the
acceleration $A$ appears in the metric function $P$ through
(\ref{specialP}) in which $x(\zeta,\bar\zeta,u)$ has to be substituted
using
  \begin{equation}
  \int{\d x\over G} =\int{\d x\over 1 -x^2 - 2mA\,x^3}
= Au -{\textstyle{1\over\sqrt2}}(\zeta+\bar\zeta).
  \label{intG}
  \end{equation}
  The cubic $G$ possess three distinct roots provided
$0<27\,m^2\,A^2\le1$ which are given by
  \begin{eqnarray}
  &&x_1 =-{1\over6mA} \,\Big[\,
1+2\cos(\varphi+{\textstyle{2\over3}}\pi)\,\Big]\>, \nonumber \\
  &&x_2 = -{1\over6mA} \,\Big[\,
1+2\cos(\varphi+{\textstyle{4\over3}}\pi)\,\Big]\>, \label{roots} \\
  &&x_3 = -{1\over6mA} \,\Big[\, 1+2\cos\varphi \,\Big]\>,
\nonumber
  \end{eqnarray}
  where
  \begin{equation}
\varphi={1\over3}\cos^{-1}(1-54m^2A^2).
  \label{varphi}
  \end{equation}
  It can be seen that \ $x_3<x_2<0<x_1$ \ and, when $mA$ is small, \
$x_1\approx1-mA$, \ $x_2\approx-1-mA$, \ and \ $x_3\approx-1/(2mA)$.

For the standard C-metric, $x$ must be in the range \ $x_2\le x\le x_1$, \
and we can express $G(x)$ in the form
  $$ G(x) =2mA(x_1-x)(x-x_2)(x-x_3). $$
  Thus, we can write
  $$ -{1\over G(x)} ={\alpha\over x-x_1} +{\beta\over x-x_2}
+{\gamma\over x-x_3}, $$
  where
  \begin{eqnarray}
  && \alpha^{-1} =2mA(x_1-x_2)(x_1-x_3), \nonumber\\
  && \beta^{-1} =-2mA(x_1-x_2)(x_2-x_3), \nonumber\\
  && \gamma^{-1} =2mA(x_1-x_3)(x_2-x_3), \nonumber
  \end{eqnarray}
  and \ $\alpha+\beta+\gamma=0$. Using (\ref{roots}), we have
  $$ {\textstyle x_1-x_2={1\over\sqrt3\,mA}\sin\varphi, \qquad
  x_1-x_3={1\over\sqrt3\,mA}\sin(\varphi+{\pi\over3}), \qquad
  x_2-x_3={1\over\sqrt3\,mA}\sin(\varphi+{2\pi\over3}), } $$
  so that $\alpha$, $\beta$ and $\gamma$ can be expressed as
  $$\alpha= {6mA\over2\cos(2\varphi-{2\pi\over3})+1}, \qquad
\beta= {-6mA\over2\cos(2\varphi-{\pi\over3})-1}, \qquad
\gamma= {6mA\over2\cos(2\varphi)+1}, $$
  where $\varphi$ is given by (\ref{varphi}). It can be seen that $\alpha$
and $\beta$ are monotonically decreasing functions of $mA$, while $\gamma$
is an increasing function and the ranges are given by
  $$ {\textstyle {1\over2}\ge\alpha>{2\over3\sqrt3}, \qquad
-{1\over2}\ge\beta>-\infty, \qquad 0\le\gamma<\infty.} $$

It is now possible to integrate (\ref{intG}) to obtain
  \begin{equation}
  (x_1-x)^\alpha (x-x_2)^\beta (x-x_3)^\gamma
=B e^{-Au}e^{(\zeta+\bar\zeta)/\sqrt2},
  \label{integral}
  \end{equation}
  where $B$ is a constant that will be determined later. This is an
implicit equation for $x$ which should be substituted into the above
expression (\ref{specialP}) for $P$. In the limit as $A\to\infty$ while
$mA$ is kept constant, we expect that the solution should approach that of
an impulsive expanding wave in a Minkowski, de~Sitter or anti-de~Sitter
background together with an associated snapping cosmic string. However, it
can be seen that the above equation for $x$ becomes degenerate in this
limit in which $x\to x_1$ for $u>0$ and $x\to x_2$ for $u<0$.
Nevertheless, it is possible to overcome this problem using the permitted
coordinate freedom \ $u\to U(u)$, \ $r\to r/\dot U$, \ $P\to P/\dot U$, \
$m\to m/\dot U^3$.

Let us first consider the case for $U>0$.
In the limit as $x\to x_1$, (\ref{integral}) implies that
  \begin{equation}
  x_1-x \approx B^{1/\alpha} (x_1-x_2)^{-\beta/\alpha}
(x_1-x_3)^{-\gamma/\alpha} \,e^{-AU/\alpha}\,
e^{(\zeta+\bar\zeta)/\sqrt2\alpha},
  \label{exp1}
  \end{equation}
  so that
  $$ P^2 \approx  C^2\, \dot U^2\,e^{AU/\alpha}\,
e^{-(\zeta+\bar\zeta)/\sqrt2\alpha}, $$
  where
  $$ C^2 = {1\over2mA} B^{-1/\alpha} (x_1-x_2)^{(\beta-\alpha)/\alpha}
(x_1-x_3)^{(\gamma-\alpha)/\alpha}. $$
  To ensure that the expression for $P$ remains finite as $A\to\infty$, we
use our coordinate freedom to set \ $C\dot U= e^{-AU/2\alpha}$ \ for $U>0$.
  Having made this transformation, the limit $A\to\infty$ for $U>0$ leads to
  $$ P^2= e^{-(\zeta+\bar\zeta)/\sqrt2\alpha}. $$

Similarly, for $U<0$, in the limit as $x\to x_2$ (\ref{integral}) implies
that
  \begin{equation}
  x-x_2 \approx B^{1/\beta} (x_1-x_2)^{-\alpha/\beta}
(x_2-x_3)^{-\gamma/\beta} \, e^{-AU/\beta}\,
e^{(\zeta+\bar\zeta)/\sqrt2\beta},
  \label{exp2}
  \end{equation}
  and
  $$ P^2 \approx D^2
  \>\dot U^2\,e^{AU/\beta}\,
e^{-(\zeta+\bar\zeta)/\sqrt2\beta}, $$
  where
  $$ D^2 = {1\over2mA} B^{-1/\beta}
(x_1-x_2)^{(\alpha-\beta)/\beta} (x_2-x_3)^{(\gamma-\beta)/\beta}. $$
  To ensure that this remains finite as $A\to\infty$, we can again use our
coordinate freedom to set \ $C\dot U= e^{-AU/2\beta}$ \ for $U>0$.
  Having made this transformation, the limit $A\to\infty$ for $U<0$ leads to
  $$ P^2= {D^2\over C^2}\, e^{-(\zeta+\bar\zeta)/\sqrt2\beta}. $$

The function $U(u)$ as specified above satisfies the conditions that $U>0$
when $u>0$ and $U<0$ when $u<0$. It is given explicitly by
  $$ U(u)=\cases{
\displaystyle{ {2\alpha\over A}\log\left(1+{A\over2\alpha C}\,u\right)}
\qquad\qquad {\rm for} \qquad u>0 \cr
\noalign{\medskip}
\displaystyle{ {2\beta\over A}
\log\left(1+{A\over2\beta C}\,u\right)}
\qquad\qquad {\rm for} \qquad u<0 } $$
  Including the additional value $U=0$ when $u=0$, the above transformation
is $C^1$, and the resulting function $U(u)$ is monotonically increasing
with $U<0$ for $u<0$ and $U>0$ for $u>0$ since $\alpha>0$ and $\beta<0$.

\begin{figure}[hpt]
\begin{center} \includegraphics[scale=0.6, trim=0 0 0 -15]{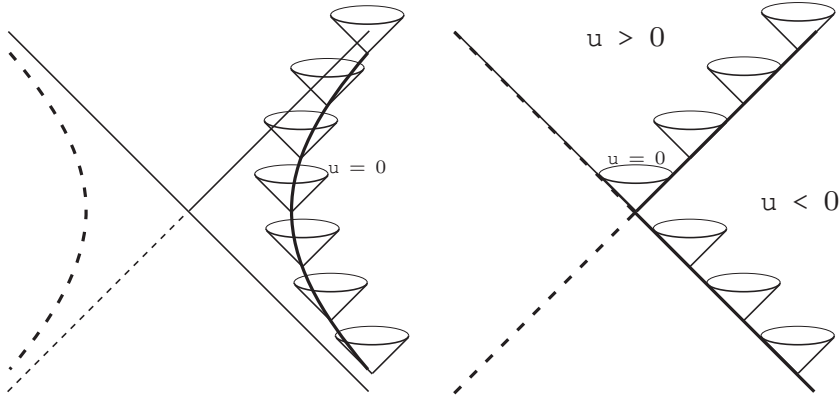}
\caption{\small In Robinson--Trautman coordinates, the null surfaces $u=$
const. define a family of null cones centred on one accelerating black
hole. These foliate half of the complete space-time (above the diagonal in
the left diagram) which is symmetric about the vertical axis. In the limit
as $A\to\infty$, the ``source'' of these cones approaches the null
asymptotes as in the diagram on the right. The null surfaces still foliate
half of the space-time, but are stacked in different ways in the two halves
$u<0$ and $u>0$. There is an obvious discontinuity in the foliation across
the impulsive wave at $u=0$. These schematic pictures apply in Minkowski,
de~Sitter and anti-de~Sitter backgrounds. }
\end{center}
\end{figure}

Before constructing the metric function $P^2(\zeta,\bar\zeta,u)$
explicitly for the whole range of $u$, however, we must observe that
according to (\ref{integral}) the two roots $x=x_1$ and $x=x_2$
correspond to the coordinate limits $\zeta\to-\infty$ and $\zeta\to\infty$
respectively. Thus, the expressions (\ref{exp1}) and (\ref{exp2})
correspond to expansions about opposite poles of the spherical wave
surfaces. In order to maintain a continuous coordinate system, it is
necessary to make the transformation $\zeta\to-\zeta$ in one of the two
regions. (The necessity of this procedure can be confirmed in the limit
when $m=0$. See also figure~2) Choosing to perform this in the region
$u>0$, we obtain
  \begin{equation}
  P^2(\zeta,\bar\zeta,u)=\cases{
  {\displaystyle P_+^2= \exp\left(
{\zeta+\bar\zeta\over\sqrt2\alpha} \right) }
  \hskip5.7pc {\rm for} \qquad u>0 \cr
\noalign{\medskip}
  {\displaystyle P_-^2= {D^2\over C^2}\, \exp\left(
{\zeta+\bar\zeta\over-\sqrt2\beta} \right) }
\qquad\qquad {\rm for} \qquad u<0 }
  \label{P+-}
  \end{equation}
  Although the discontinuity at $u=0$ is consistent with the presence of an
impulsive wave, this is not yet expressed in the form given above for an
impulsive Robinson--Trautman solution. However, we can now make use of the
remaining coordinate freedom that $\zeta\to\tilde\zeta=h(\zeta)$ under which
$P^2\to\tilde P^2=h'\bar h'P^2$. \ In particular, we can make the
specific choice \ $h=\sqrt2\alpha\exp\left(-\zeta/\sqrt2\alpha\right)$ \
so that
  $$ \tilde P_+^2=1, \qquad
  \tilde P_-^2={D^2\over C^2}
\left({\tilde\zeta\bar{\tilde\zeta}\over2\alpha^2}\right)
^{1+(\alpha/\beta)}. $$
  Dropping the tildes, we can then put \ $g=-\alpha/\beta\in(0,1]$ \ and
choose the constant $B$ such that \
$2\beta^2(2\alpha^2)^{\alpha/\beta}C^2=D^2$, \  i.e. \
$B^{(1/\alpha)-(1/\beta)} =(2\alpha^2)^{1+\alpha/\beta}
(\alpha/\gamma)^{\alpha/\beta} (-\gamma/\beta)^{\beta/\alpha}$. \
We thus finally obtain 
  \begin{equation}
  P^2(\zeta,\bar\zeta,u)=\cases{
  {\displaystyle \qquad 1 }
  \hskip6.8pc {\rm for} \qquad u>0 \cr
\noalign{\medskip}
  {\displaystyle \quad g^{-2}\,(
\zeta\bar\zeta)^{1-g}
\qquad\qquad {\rm for} \qquad u<0 } }
  \label{Pfinal}
  \end{equation}
  which is exactly the form given above for an impulsive Robinson--Trautman
wave (\ref{functions}) with $\epsilon=0$ and $g$ given by~(\ref{g}). In
fact, it can be seen from figure~2 that the foliation of the space-time in
either regions $u<0$ or $u>0$ by the null cones $u=$~const. is only
consistent with the parametrization of the Robinson--Trautman family of
solutions with $\epsilon=0$ as described in~\cite{GrPoDo02}. In this
parametrization, the solution (\ref{Pfinal}) describes an impulsive
spherical wave generated by the snapping of a cosmic string with deficit
angle $2\pi(1-g)$ in a Minkowski, de~Sitter or anti-de~Sitter background
according to the value of $\Lambda$. It reduces to the corresponding
complete background with no string and no impulse when $mA=0$, in which
case \ $\alpha={1\over2}=-\beta$, \ $g=1$, \ and \ $B=1$ \ so that
$P=1$ everywhere. It also agrees with the null limit of the C-metric
\cite{PodGri01b} that was previously obtained only for the Minkowski
background.

\section{Concluding remarks}

The form of the C-metric considered above describes a pair of black holes
which accelerate away from each other in a Minkowski, de~Sitter or
anti-de~Sitter universe under the action of cosmic strings which pull them
apart. By adopting a Robinson--Trautman coordinate system, we have shown
that this type~D space-time leads to a solution which describes a
snapping cosmic string in any of these backgrounds in the limit in which
the acceleration $A$ becomes unbounded while the mass $m$ is scaled to zero
such that $mA$ is kept constant. This procedure leads to a metric form
which is discontinuous over the spherical wavefront $u=0$ generated by the
snapping string. Such a discontinuity seems to be inevitable in this
coordinate system due to the difference in the foliation of the
space-time in the two regions $u<0$ and $u>0$ as illustrated in figure~2.
Nevertheless, the interpretation given above is justified because the
limit obtained is identical to that which arises as the impulsive limit of
a specific family of type~N sandwich waves of Robinson--Trautman type.

In section~4, we considered such a special family of Robinson--Trautman
type~N space-times which describe expanding sandwich waves with spherical
wavefronts again in Minkowski, de~Sitter or anti-de~Sitter
backgrounds~\cite{GrPoDo02}. These were given for all possible values of
the parameter~$\epsilon$ which determines the foliation of the space-time.
The family of solutions considered were previously shown to contain a pair
of cosmic strings ahead of the wave, but have no string behind it. Apart
from the presence of a string-like singularity within the wave region,
these solutions are regular and well behaved everywhere. It is then very
reasonable to investigate these solutions in the limit as the duration of
the wave becomes arbitrarily small. Such a limit leads unambiguously to an
expression for a spherical impulsive gravitational wave, but the metric
again becomes discontinuous and is exactly the null limit of the C-metric
described above.

In section~3, we used the ``cut and paste'' method as another way to
construct spherical impulsive waves. Here, we presented a solution in
which the metric is explicitly continuous and also valid for all possible
values of the cosmological constant and the parameter~$\epsilon$. These
provide an alternative form of the above solutions for a snapping string
together with an expanding impulse. We have indicated how this form of the
metric is related to the 5-dimensional representation of the background
de~Sitter and anti-de~Sitter universes.

When $\Lambda=0$, the null limit of the C-metric was previously derived in
an explicitly continuous form~\cite{PodGri01b}. This could be achieved
because a global coordinate system adapted to the boost-rotation symmetry
was available. However, such coordinates are still unknown for the C-metric
with a non-zero cosmological constant. Nevertheless, we have been able to
derive the solution for a snapping cosmic string generating an expanding
impulsive spherical gravitational wave in a de~Sitter and anti-de~Sitter
universe by three alternative methods. These different approaches
complement each other and lead to a fairly complete understanding of these
space-times.

\section*{Acknowledgements}

The work of JP was partly supported by the grants GA\v{C}R-202/02/0735,
GAUK~166/2003 and the London Mathematical Society.

\end{document}